\documentclass[aps,prb,twocolumn]{revtex4-2}
\usepackage{graphicx}
\usepackage{amsmath}
\usepackage{amssymb}
\usepackage{braket}
\usepackage{hyperref}
\usepackage{subfigure}
\usepackage[usenames,dvipsnames]{xcolor}
\usepackage{graphicx}
\usepackage[mathscr]{euscript}
\usepackage{comment}
\usepackage{caption}

\begin{document}

\preprint{AIP/123-QED}

\title[]{Quasi-Periodically Driven Quantum Ising Chains}
\author{Dhruvil Doshi}
 \altaffiliation[]{Physics Department, IIT Kanpur}


\begin{abstract}
Understanding the equilibration of isolated quantum systems under unitary dynamics is an interesting topic.
In this paper we look at the early time behaviour of periodically and quasi-periodically driven Transverse field Ising chains when and their corresponding dynamical free energies. We study the system under different frequencies and observe how the system evolves with changes in the field amplitudes in both types of oscillations.
\end{abstract}

\maketitle

\section{\label{sec:level1} Introduction (DQPT)}
Dynamical Quantum Phase transitions (DQPT) mean phase transitions observed under time evolution. Similar to the classical analogue where we subjected temperature as the parameter, in DQPTs, the corresponding parameter is the complex time.[1]
In statistical mechanics, we treat the Partition function (Z) as the primary entity, while in DQPT we define the Loschmidt Amplitude $G(t)$ as our primary object. It is defined as 
$G(t) = \braket{\Psi_{o} | \Psi_{o}(t)}$. This gives us the deviation of the time evolved state form the initial state.\\
The Loschmidt echo (corresponding probability) is defined as $L(t) = |G(t)|^{2}$.\\
We further define a rate function $\lambda(t) = - lim_{N \to \infty} \frac{1}{N} log(L(t))$.
DQPTs are defined as the non-analytical evolution of the rate function(dynamical free energy density).
One important note is that we can analytically find the critical times $t_c$ from the Loschmidt echo i.e when $L(t_c) = 0$.\\  
Transverse field Ising Chain is the quantum analogue of the classical Ising chain. In this paper we will look at different dynamics of this system when it undergoes quenching, periodic driving and quasi-periodic driving.

Its Hamiltonian is defined as (for $J>0$):
\begin{equation}
 H_{1,2} = -J\sum^{N}_{j=1} \sigma^{z}_{j} \sigma^{z}_{j+1} + h_{1,2} \sum^{N}_{j=1} \sigma^{x}_{j}.
\end{equation}

\section{\label{sec:level2}Transverse Field Ising Model}

Fermionic Formulation - The Jordan-Wigner Transformation helps us map the spin operators to spinless fermionic creation and annihilation operators, which aids us in solving the Ising model exactly by diagonalizing in the fermionic basis. The primary idea stems from the fact that spin 1/2 systems are akin to spinless fermions.[3]\\
The $c_{j}$ and $c_{j}^{\dagger}$ operators are the fermionic creation and annihilation operators.\\
The transformation is essentially :-
\begin{subequations}\label{eq_jw}
\begin{gather}
\sigma^{x}_{j} = K_{j} (c^{\dagger}_{j} + c_{j}),\\
\sigma^{y}_{j} = K_{j} i(c^{\dagger}_{j} - c_{j}),\\
\sigma^{z}_{j} = (1 - 2n_{j}).\\
\mbox{Where }
K_{j} = \prod_{l=1}^{j-1} (1 - 2n_{l}).
\end{gather}
\end{subequations}
$K_{j}$ basically takes the value of +1 or -1 depending on the number of fermions present before site j.\\
The fermionic operators in the k-space (momentum space) are related to the real space by the following transformation for systems with periodic boundary condition :-
\begin{equation}\label{k_trans}
c_k = \frac{1}{\sqrt{L}} \sum^{L}_{j=1} e^{-ikj}c_{j}.
\end{equation}
By writing the Hamiltonian in terms of the fermionic operators, we can see that the particle number is not conserved but the parity of particle number is conserved.
Using the above transformation equations (Eq.~\eqref{eq_jw}) and converting the fermionic operators in the momentum space (Eq.~\eqref{k_trans})(joining the +k and -k terms), we can write the Hamiltonian as a summation of 2X2 matrices in the momentum modes. This helps us diagonalize the Hamiltonian and find the eigenvalues and eigenvectors.\\
\begin{subequations}\label{Hk}
\begin{gather}
H = \sum^{K}_{k} H_{k},\\
\label{Hk2}
H_{k} = (c_k^{\dagger}, c_{-k})
\begin{pmatrix}
2(h-Jcosk) & -2iJsink \\
2iJsink & -2(h-Jcosk) 
\end{pmatrix}
\begin{pmatrix}
c_k\\
c_{-k}^{\dagger}
\end{pmatrix},\\
K = \{ \frac{2n\pi}{L}  where\: n = 1,..., \frac{L}{2} -1 \}.
\end{gather}
\end{subequations}
The eigenvalues of the Hamiltonian will be degenerate (energy gap = 0) at the quantum critical point which is obtained when $|h_{c}| = J$. For $|h_{c}| < J$ the model behaves like a ferromagnetic phase (interaction term dominates) while for $|h_{c}| > J$ the model is in paramagnetic phase (external transverse field dominates).\\
Eq.~\eqref{Hk} is the primary equation we use for most of the numerical analysis since it is easily diagonalizable and exactly solved.\\
\textbf{Note:} All above equations were  derived for the periodic boundary condition. 
\section{\label{sec:level3}Quenching}

Quenching means we instantaneously change a parameter. In the Transverse Field Ising Hamiltonian (TFIH), we quench the transverse external field i.e. suddenly set the h (external field) from the initial value $h_i$ to a new $h_{f}$.
\\
Thus the $H_{f} = -J\sum^{N}_{j=1} \sigma^{z}_{j} \sigma^{z}_{j+1} + h_{f} \sum^{N}_{j=1}$.\\

The following plot(left) displays the non-analyticity in the free energy when we quench the transverse field.\\
These DQPTs are observed only when the quench crosses the critical point i.e when $h_i>J$  and $h_f<J$.
\begin{figure}[h]
 \centering
\includegraphics[width = 70mm]{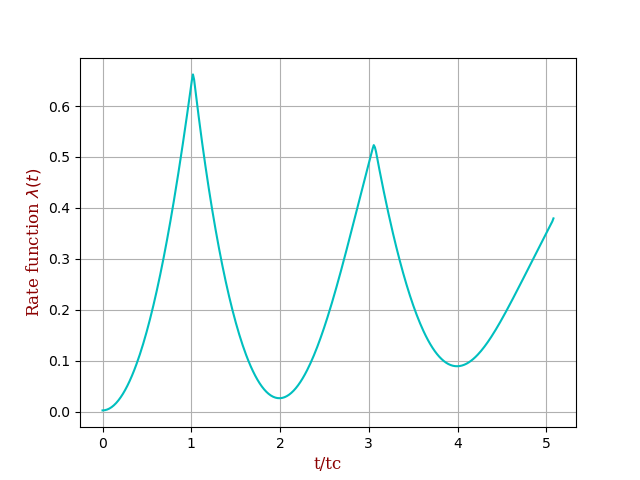} 
\includegraphics[width = 70mm]{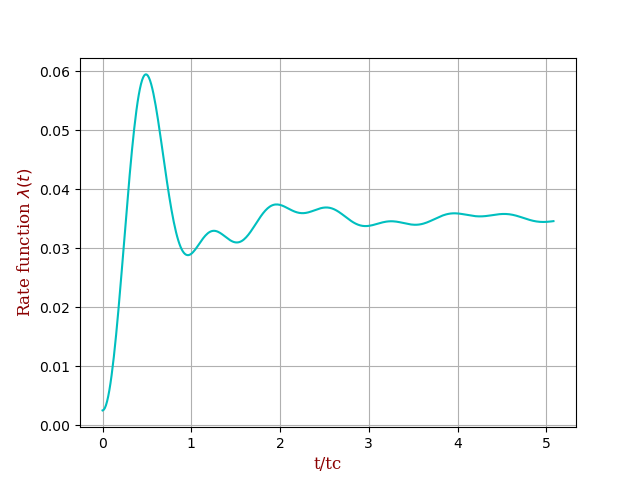}
 \caption{Top: Quench from $h=10J$ to $h=0.2J$
 Bottom: Quench from $h=10J$ to $h= 2J$}
 \label{fig5}
\end{figure}

Clearly, we can see that the DQPTs occur at odd multiples of the critical time in .\\
Further we can observe from the Fig 1b that if the transverse field does not cross the critical value ($h = J$) then no DQPTs are observed.\\

\section{\label{sec:level4}Periodic Driving}
Periodic driving is when we oscillate the external field between two values in a periodic manner. In one stroboscopic time period ($T$), the Hamiltonian is as follows ($h_{1}$ and $h_{2}$ are transverse external field magnitudes): \\
\begin{equation}
H(t) = 
\left\{
    \begin{array}{lr}
        -J\sum^{N}_{j=1} \sigma^{z}_{j} \sigma^{z}_{j+1} + h_{1} \sum^{N}_{j=1} \sigma^{x}_{j}, & \text{if } t \leq T/2\\
        -J\sum^{N}_{j=1} \sigma^{z}_{j} \sigma^{z}_{j+1} + h_{2} \sum^{N}_{j=1} \sigma^{x}_{j}. & \text{if } t > T/2
    \end{array}
\right\}
\end{equation}
This periodic driving keeps on repeating. The driving frequency ($\omega= 2\pi/T $) is an essential parameter which affects the dynamics of the systems evolution.
We will look at the exact numerical plots for different amplitude values of the external field as well as for different frequency values.\\

We can see that for high frequency oscillations, the results appear identical to what we see if the system underwent a quench from  $h_i$ to $h_f = h_{eff} = \frac{(h_{1} + h_{2})}{2} $.\\
We can further verify this by checking that if we set $h_1$ and $h_2$ such that the $h_{eff}$ does not cross the critical point, then no DQPTs are observed. \\
\begin{figure}[h]
 \centering
\includegraphics[width = 70mm]{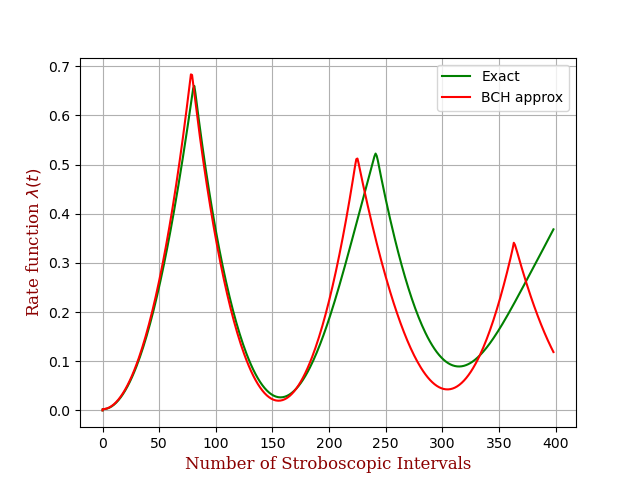}
\includegraphics[width = 70mm]{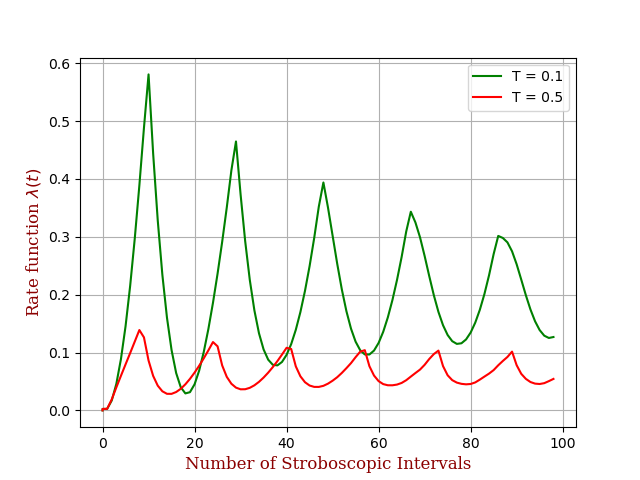} 
 \caption{Top: High frequency periodic oscillations from $h=10J$ to $h=-9.6J$ \newline
 Bottom: Low frequency periodic oscillations on increasing time periods}
 \label{fig5}
\end{figure}
The above results can be explained analytically when we can approximate the equation $ \ket {\psi (nT)} = U_F ^{n}\ket{\psi (T)}$ by using the BCH expansion and neglecting the terms in the exponent with higher orders of T. \\
$H_f (k) = \frac{1}{2}(H_1 ^{k} + H_2 ^{k}) + \frac{T}{8i} [H_2 ^{k}, H_1^{k} ] + O(T^{2})$\\
At low frequencies, we notice that the DQPTs eventually disappear when the the time period is comparable to the order of the interaction term (J). We cannot accurately comment as to which set of parameters and time periods will result in DQPTs at low frequencies, since approximating it to the quenching conditions is not justified anymore.

\vspace{-5mm}
\section{\label{sec:level5}Quasi-Periodic Driving}
Quasi-Periodicity means that the system evolves as in an irregular (unpredictable manner). In this paper we focused on Fibonacci driving.[2]\\
The best way to comprehend this evolution is from the table below.
\vspace{2mm}\\
\begin{tabular}{ |p{3.7cm}|p{3cm}|p{2cm}|  }
 \hline
 \multicolumn{3}{|c|}{Fibonacci Driving} \\
 \hline
 Fibonacci Sequence & Number of stroboscopic periods (N)& Number of 1 and 2\\
 \hline
 1   & 1    & 1, 0\\
 12 &  2  & 1, 1\\
 121 & 3 & 2, 1\\
 12112    & 5 & 3, 2\\
 12112121 &  8  & 5, 3\\
 1211212112112 & 13  & 8, 5\\
 121121211211212112121 & 21  & 13, 8\\
 \hline
\end{tabular}
\vspace{1mm}
\\
Using this above sequence, I have computed the exact numerical solutions for both high and low frequencies.

Now we will define 3 well established parameters - $\alpha$, $\beta$, $\delta$ for this Fibonacci driving to analytically approximate the equation for high frequencies. $\alpha(N)$ and $\beta(N)$ are the number of times 1 and 2 occur in N stroboscopic instances respectively. The time evolution operator is the multiplication of N matrices of $e^{-iH_{1}T}$ and $e^{-iH_{2}T}$ in the Fibonacci sequence. We can now expand this exponential matrix multiplication using the BCH expansion. We expand terms till the first order of T. $\delta(N)$ is the number of commutators of $H_1$ and $H_2$ present in this expansion.\\

Using the BCH expansion, we can approximate the unitary time evolution operator as $U_f (N) = e^{iNT H_f}$ where\\ $H_f (N) = \frac{\alpha (2N)}{2N} H_1 ^{k} + \frac{\beta (2N)}{2N} H_2 ^{k} + \frac{T}{4i} [H_1 ^{k}, H_2^{k} ] + O(T^{2}).$\\
\begin{figure}[h]
 \centering
\includegraphics[width = 70mm]{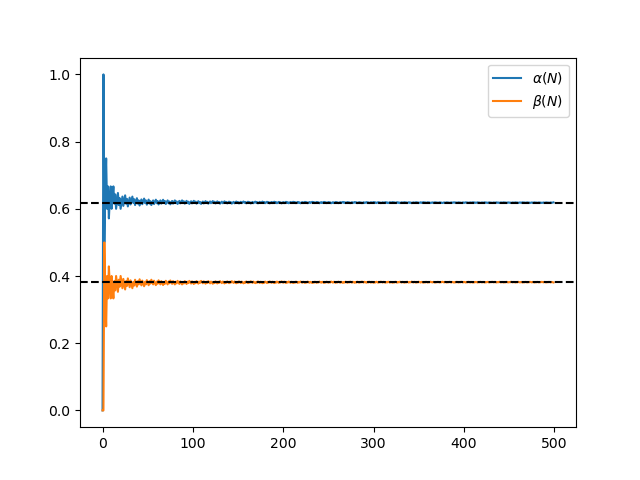} 
\includegraphics[width = 70mm]{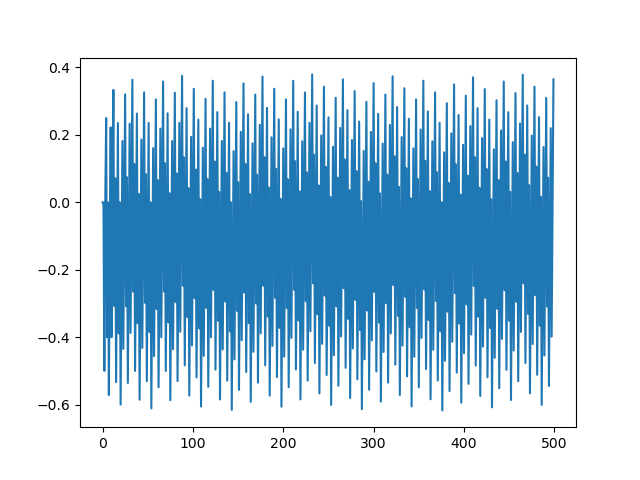}  
 \caption{Top: $\alpha/(N)$ and $\beta/(N)$ Bottom: $\delta/(N)$}
 \label{fig5}
\end{figure} 

We can see that for high frequency oscillations, the results appear identical to what we see if the system underwent a quench from  $h_i$ to $h_f = h_{eff} = \frac{\alpha }{N}h_{1} + \frac{\beta}{N}h_{2} $ for very high freq (ignoring even T terms).\\
We can further verify this by checking that if we set $h_1$ and $h_2$ such that the $h_{eff}$ does not cross the critical point, then no DQPTs are observed.

\begin{figure}[h]
 \centering
 \includegraphics[width = 70mm]{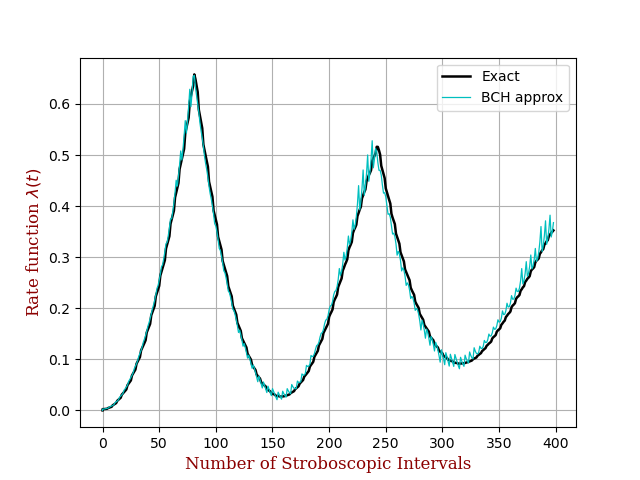}  
 \caption{High frequency quasi periodic oscillations from $h=10J$ to $h=-15.65J$ }
 \label{fig5}
\end{figure}
\begin{figure}[h]
 \centering
\includegraphics[width = 70mm]{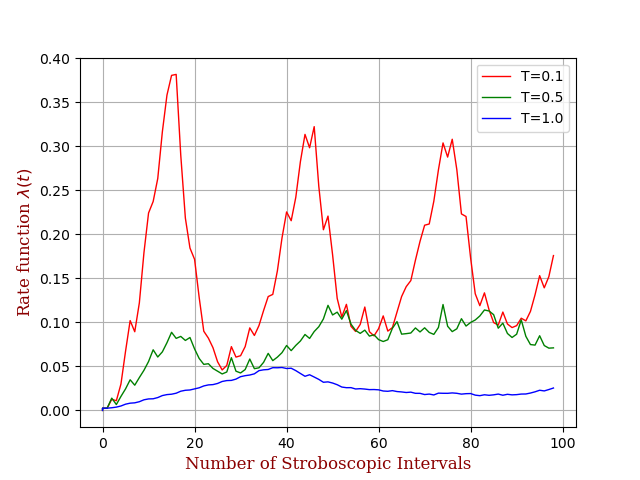} 
 \caption{ Low frequency quasi periodic oscillations on increasing time periods}
 \label{fig5}
\end{figure}

In quasi periodic driving at low frequencies($T$ comparable to order of $J$), we observe that the existence of DQPTs cannot be accurately estimated since approximating it to the quenching conditions is not justified anymore.

\newpage

\section{\label{sec:level6}Results}
From all the above analytical and numerical computations, we proved that the time analysis of the transverse field Ising model for different drivings can we approximated to a quench with transverse field of $h_{eff}$ in the high frequency limit ($w>>J$) where $h_{eff}$ is based on the driving parameters.\\
While the analysis for periodic driving is already well-established, quasi-periodic (Fibonacci driving) is a new result which can also be subjected to these estimations and analysis.\\

\vspace{-10mm}
\section{Acknowledgements}
This text is a report of the undergraduate semester project I had taken up under the supervision of Professor Amit Dutta. I would like to thank him for his advice throughout the project. I am indebted to him for his guidance and support. I am also grateful to Sourav Bhattacharjee and Saikat Mondal for their invaluable assistance and feedback.

\section{Bibliography}
1) Heyl, M. (2018). Dynamical quantum phase transitions: a review. Reports on Progress in Physics, 81(5), 054001.\\
2) Maity, S., Bhattacharya, U., Dutta, A., Sen, D. (2019). Fibonacci steady states in a driven integrable quantum system. Physical Review B, 99(2), 020306.\\
3) Mbeng, G. B., Russomanno, A., Santoro, G. E. (2020). The quantum Ising chain for beginners. arXiv preprint arXiv:2009.09208.\\
4) Bhattacharjee, S., Bandyopadhyay, S., Dutta, A. (2022). Quasi-localization dynamics in a Fibonacci quantum rotor.\\
\end{document}